\documentclass[aps,prl,twocolumn,showpacs,superscriptaddress,nofootinbib,preprintnumbers]{revtex4-2}
\usepackage[utf8]{inputenc}
\usepackage{amssymb}
\usepackage{hhline}
\usepackage{amsmath}
\usepackage{mathtools}
\usepackage[dvipsnames]{xcolor}
\usepackage{multirow,tabularx}
\usepackage{graphicx}
\usepackage{graphicx}
\usepackage{ulem}

\usepackage{xspace}
\usepackage{xstring}
\usepackage{parskip}
\allowdisplaybreaks
\usepackage{braket}

\newcommand{\vo}{$v_0$\xspace}
\newcommand{\vesc}{$v_\mathrm{esc}$\xspace}

\newrobustcmd{\pea}[1]{%
	\emph{#1}\textbf{\ \ \ ---}
}

\usepackage[colorlinks=true,citecolor=blue,urlcolor=blue]{hyperref}
\hypersetup{colorlinks=true,citecolor=romared,linkcolor=romared,urlcolor=romared}
\definecolor{romared}{RGB}{142,0,28}

\definecolor{tabblue}{RGB}{31, 119, 180}
\definecolor{darkblue}{RGB}{0, 0, 120}
\definecolor{tabred}{RGB}{214, 39, 40}
\definecolor{tabgreen}{RGB}{44, 160, 44}
\definecolor{tabgray}{RGB}{100, 100, 100}
\begin{document}

\title{Confronting the Higgsino Interpretation of the LZ Event with Astrophysical Uncertainties and Constraints from Solar Capture }


\author{Aditya Ghosh}
\thanks{Corresponding author}
\email{n01546521@unf.edu}
\affiliation{
Department of Physics and Astronomy, University of North Florida, Jacksonville, Florida 32224, USA
}

\author{Ilumi Chavez}

\affiliation{
Department of Physics and Astronomy, University of North Florida, Jacksonville, Florida 32224, USA
}

\author{Chris Kelso}
\affiliation{
Department of Physics and Astronomy, University of North Florida, Jacksonville, Florida 32224, USA
}

\begin{abstract}
We investigate whether astrophysical uncertainties in the local dark matter distribution can alleviate tensions between the higgsino interpretation of the LZ event and constraints from solar-capture neutrinos. We calculate the expected LZ signal over a wide range of Standard Halo Model parameters and using $N$-body hydrodynamical simulations of Milky Way-like galaxies that include the influence of the Large Magellanic Cloud. For the Standard Halo Model, the requirement $\delta \gtrsim 566~\mathrm{keV}$ implies $v_0 \gtrsim 240~\mathrm{km\,s^{-1}}$, independent of the Galactic escape speed. In the simulated halo, we find that $\delta = 514~\mathrm{keV}$ is the largest mass splitting capable of producing one event in LZ's $200~\mathrm{keV}-300~\mathrm{keV}$ recoil-energy range. We conclude that astrophysical uncertainties are insufficient to reconcile the higgsino interpretation of the LZ event with solar-capture neutrino constraints.

\end{abstract}

\maketitle

\textit{\textbf{Introduction}} -- Despite decades of intense investigation, the nature of dark matter remains one of the most compelling mysteries of fundamental physics.  The LUX-ZEPLIN (LZ) Collaboration has recently presented results from a search for dark matter (DM) interactions in an extended nuclear-recoil energy window up to approximately 300\,{\rm keV}~\cite{LZ:2026axp}.  The LZ Collaboration reported an event consistent with a nuclear recoil of $248\pm23\,({\rm stat})\pm23\,({\rm sys})\,{\rm keV}$ in this region, where the expected background rate is very low. The event is difficult to accommodate with conventional elastic, spin-independent DM scattering, which generally predicts a spectrum concentrated at lower recoil energies, but could arise more naturally from momentum-dependent or inelastic interactions~\cite{Fan:2026kxx,Freese:2026sga,Wu:2026nhi,Yin:2026jnn,Su:2026rwz,Yamashita:2026ump,DiMauro:2026ldr}. Across the models considered by LZ, the largest local significance was $3.4\sigma$, which was reduced to a global significance of $2.6\sigma$ after accounting for the look-elsewhere effect. 

An exciting possibility for the source of the signal is inelastic scattering of a supersymmetric higgsino of mass $\sim1.1\,\mathrm{TeV}$ with a xenon nucleus in the detector. The correct dark matter relic density for a 1.1 TeV higgsino is achieved through annihilation and coannihilation, primarily into gauge bosons, with another neutralino and chargino of similar mass during thermal freeze-out in the early universe.  A similar model was examined by the LZ Collaboration as a possible explanation for the event in their original analysis~\cite{LZ:2026axp}.  The higgsino interpretation of the LZ event has subsequently been explored and tested further by several other groups \cite{Freese:2026sga, Fan:2026kxx, DiMauro:2026ldr, RoddEtAl:2026LZSideband, Pospelov:2026ewn, BoseEtAl:2026HiggsinoNeutrinos,DiMauroShaikh:2026SolarCapture, NguyenLindenHooper:2026SolarNeutrinoLZ, Langhoff:2026HeavyHiggsino, WuZhangZhu:2026TeVHiggsino, DuWang:2026HiggsinoLZ, Cheung:2026LZHiggsinoCollider, Yin:2026PQSUSYLZ}.

Additionally, soon after LZ reported their results, Ref.~\cite{Pospelov:2026ewn} presented an analysis of solar capture and subsequent higgsino annihilation through $W^+W^-$ and $ZZ$ channels.    The non-observation by the IceCube experiment of a flux of high-energy neutrinos from the annihilation of these captured DM particles in the Sun places strong constraints on the mass splitting, $\delta$.  Repeated up-scattering with uranium through tree-level $Z$ exchange requires $\delta \lesssim 506\,{\rm keV}$, while a proposed loop-induced elastic scattering of the captured DM population strengthens the bound to $\delta \lesssim 566\,{\rm keV}$.  Subsequent analyses using IceCube and/or Super-Kamiokande confirm these results~\cite{Bose:2026ndd,Nguyen:2026lui}.

In this Letter, we quantify how astrophysical uncertainties in the dark matter distribution affect the higgsino interpretation of the LZ event, with particular emphasis on constraints from solar-capture neutrinos.

\textit{\textbf{Expected Signal}} -- The spectrum of DM scattering events at a detector such as LZ is given by
\begin{equation}
\frac{dR}{dE_R} = N_T \frac{\rho_{\chi}}{m_{\chi}} \int_{|\vec{v}|>v_{\rm min}} d^3v \, v \, f_E(\vec{v}) \, \frac{d\sigma}{dE_R},
\label{eq:dRdEr}
\end{equation}
where $N_T$ is the number of target nuclei, $m_\chi$ is the DM mass, $\rho_{\chi}$ is the local DM density,   $f_E(\vec{v})$ is the distribution function of DM particle velocities, $\vec{v}$, in the frame of the Earth, and $\frac{d\sigma}{dE_R}$ is the differential scattering cross section.  The minimum speed required to produce a nuclear recoil of energy $E_r$ for a nuclear target of mass $m_T$ and reduced mass $\mu_T=m_\chi m_T/(m_\chi+m_T)$, is given by
\begin{equation}
v_{min}(E_R,\delta)
=
\frac{1}{\sqrt{2m_TE_R}}\left(\frac{m_TE_R}{\mu_T}+\delta\right),
\label{eq:vmin}
\end{equation}
where $\delta$ is the energy threshold for inelastic dark matter scattering to occur, which is equal to the mass difference between the two lightest neutralinos in this higgsino model.  To obey the solar capture neutrino bounds of $\delta\gtrsim566\,\mathrm{keV}$ (or $\delta\gtrsim506\,\mathrm{keV}$), a thermal higgsino of mass $m_\chi\sim1.1\,\mathrm{TeV}$ inelastically scattering with a xenon nucleus to produce of recoil of $\sim248\,\mathrm{keV}$ requires Earth/lab frame speeds of $v_{\rm min}\,\sim 1025\,\mathrm{km~s^{-1}}$ (or $v_{\rm min}\,\sim 952,\mathrm{km~s^{-1}}$).

The velocity distribution of DM is typically parameterized in terms of a Maxwell-Boltzmann distribution, truncated above the local escape speed of the Galaxy, $v_{\rm esc}$:
\begin{align}
f_{\rm gal}(\vec{u})
&=
\frac{
e^{-u^2/v_0^2}\Theta(v_\mathrm{esc}-u)
}{
\mathcal N\pi^{3/2}v_0^3
},
\\
\mathcal N
&=
\operatorname{erf}(z)
-
\frac{2z}{\sqrt{\pi}}e^{-z^2},
\qquad
z=v_\mathrm{esc}/v_0.
\label{eq:shm}
\end{align}
where $\vec{u}$ is the velocity of the dark matter particle in the galaxy frame. The velocity distribution seen by an observer in the Earth frame will be ``boosted'' by the velocity of the Earth relative to the galaxy, $\vec{v}_E$ via the relationship $f_E(\vec{v})=f_\mathrm{gal}(\vec{v}+\vec{v}_E)$.  This velocity distribution is commonly used by the direct detection community and is referred to as the Standard Halo Model (SHM).  

The SHM has three parameters: $\rho_\chi, \vec{v}_E,v_\mathrm{esc}$, with $\vec{v}_E$ typically referred to as the local standard of rest and normally denoted as $\vec{v}_0$.  We present as the measured values for these parameters in our work, the values adopted be many researchers and experimental collaborations within the field recommended in Ref.~\cite{Baxter:2021pqo}. The values for $\rho_\chi$ vary significantly depending on the technique, but typically fall in the range of 0.2 to $0.6\,\mathrm{GeV~cm^{-3}}$, with $0.3~\mathrm{GeV~cm^{-3}}$ the canonical value recommended.  For completeness, we show results using both $0.3~\mathrm{GeV~cm^{-3}}$ and $0.45~\mathrm{GeV~cm^{-3}}$.  The recommended value for the escape speed is $544~\mathrm{km~s^{-1}}$ for consistency with historical work, but Ref.~\cite{Baxter:2021pqo} presents measurements with 90\% confidence levels that extend down to $477~\mathrm{km~s^{-1}}$ and up to $706~\mathrm{km~s^{-1}}$.  The recommended value for $v_0$ is $238~\mathrm{km~s^{-1}}$, with $2\sigma$ uncertainties among all measurements in Ref.~\cite{Baxter:2021pqo} extending from $180~\mathrm{km~s^{-1}}$  up to $256~\mathrm{km~s^{-1}}$.

Our analysis is focused solely on the expected signals at LZ from the thermal higgsino under different astrophysical assumptions. We have accounted for LZ's efficiency in the region of interest through
\begin{equation}
N_i=MT\int_{\Delta E_{i}}dE_R\,\epsilon(E_R)\frac{dR}{dE_R},
\label{eq:counts}
\end{equation}
where $N_i$ represents the number of events in the $i$-th bin, $\Delta E_{i}$ is the energy range of that bin, $MT$ the exposure of the detector (2.84 tonne-years),  $\epsilon(E_R)$ is LZ's efficiency (digitized from Ref.~\cite{LZ:2026axp}), and $\frac{dR}{dE_R}$ given by Eq.~\ref{eq:dRdEr}.  We utilize the WimPyDD~\cite{WimPyDD2021} code to perform the calculations and include the factor of 4 in the cross section that was pointed out in Ref.~\cite{Pospelov:2026ewn}.

With only one single event detected by LZ, a full spectral analysis of the signal is impossible.  We still include some level of spectral information by considering a two-bin analysis, with a low energy range from $i=1\,\mathrm{keV}-200\,\mathrm{keV}$, and a high energy range from $i=200\,\mathrm{keV}-300\,\mathrm{keV}$.  While the choice of $200\,\mathrm{keV}$ is somewhat arbitrary, it is a value that sits comfortably outside the energy of the measured event and its quoted $1\sigma$ uncertainties. This treatment is slightly different from many studies that have subsequently been conducted after the announcement of the LZ result that examined scenarios that produce 1 event in the full LZ region of interest.  The two high-energy rates we investigate are $N_{200-300}=1$ and $N_{200-300}=-\ln(0.95)=0.051293\ldots,$ for which a signal-only Poisson process has a $5\%$ probability of producing at least one event. This second benchmark provides a scenario where the expected rate is below the nominal detection threshold, but might have an upward fluctuation leading to the LZ event.

\begin{figure*}
    \centering

    \includegraphics[
        width=0.96\textwidth
    ]{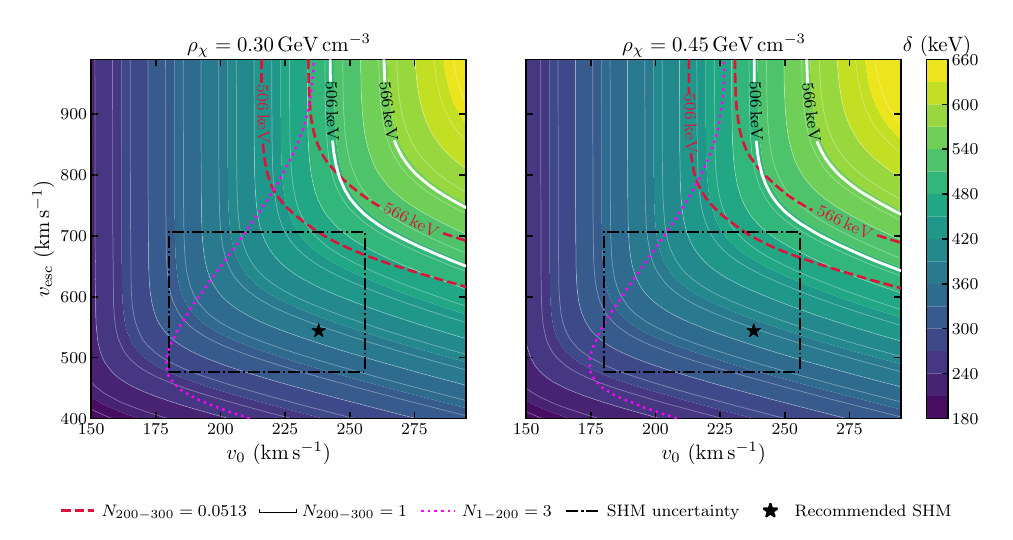}

    \caption{
    A scan over the \vo and \vesc parameters of the SHM showing contours of the mass splitting, $\delta$, required for a $1.1~{\rm TeV}$ thermal higgsino to produce one expected event in LZ's $200~\mathrm{keV}-300~\mathrm{keV}$ recoil window (after accounting for efficiency). The local dark matter density is assumed to be $ 0.30~\mathrm{GeV~cm^{-3}}$ ($0.45~\mathrm{GeV~cm^{-3}}$) in the left (right) frame.  Gray contours show the $N_{200-300}=1$ event and red dashed contours show the $N_{200-300}=0.0513$ events (the rate with a 5\% chance of an upward fluctuation producing 1 event).  The $\delta=506~\mathrm{keV}$ and $\delta=566~{\rm keV}$ contours are highlighted as the bounds from Ref.~\cite{Pospelov:2026ewn}. The black star marks the recommended SHM values, the horizontal dot-dashed lines show the full range of 90\% confidence intervals from all measurements of \vesc, and the vertical dot-dashed lines show the full range of $2\sigma$ uncertainties among all measurements of \vo, with all of these measured values presented in Ref.~\cite{Baxter:2021pqo}.  The dotted magenta curve shows the contour where 3 signal events are predicted at LZ in the lower energy range from $1-200\,\mathrm{keV}$ (where 0 signal events were detected), with points to the left of this contour being disfavored.
    }

    \label{fig:scan}

\end{figure*}

\textit{\textbf{Results for SHM}} -- Fig.~\ref{fig:scan} presents a scan in the $v_0,v_\mathrm{esc}$ plane of the SHM for 1.1 TeV thermal higgsino that produces the correct relic density assuming a local dark matter density of $\rho_\chi=0.3\,\mathrm{GeV\,cm}^3$ ($0.45\,\mathrm{GeV\,cm}^3$) in the left (right) frame.   The heat map and gray contours display the $\delta$ required to produce one expected LZ event in the $200~{\rm keV}-300~{\rm keV}$ recoil window (accounting for the LZ efficiency), with the relevant contours of 506\,keV and 566\,keV highlighted in white.  The dotted magenta curve shows the contour where 3 signal events are predicted at LZ in the lower energy range from $1-200\,\mathrm{keV}$ (where LZ detected 0 signal events).  This rate of $N_{1-200}=3$ has a 5\% probability of a downward fluctuation producing 0 signal events, with points to the left of this contour being disfavored as they would have $N_{1-200}>3$.  The recommended values for these parameters (black star), along with their uncertainty values discussed previously (black dot-dashed) are also shown.  We have performed the same scans for the $N_{200-300}=0.0513$ scenario, but only display the 506 keV and 566 keV contours in the figure as red dashed lines for the sake of brevity and ease of comparison between the two scenarios.

We find that the $\delta\gtrsim566\,\mathrm{keV}$ bound would require SHM parameter values that are well beyond any of the uncertainty ranges quoted for \vo and \vesc, even in the very optimistic scenario of $\rho_\chi=0.45\,\mathrm{GeV}$ and an upwards fluctuation of the $N_{200-300}=0.0513$ rate.   If the bound can be relaxed to $\delta\gtrsim506\,\mathrm{keV}$, we do find viable values of the SHM that can produce the LZ  signal if both \vo and \vesc are pushed to the upper limits of their uncertainties, even in the $\rho_\chi=0.30\,\mathrm{GeV}$ benchmark.  

One interesting feature of the contour plot is that for a given value of \vo, there is a point from which increasing $v_\mathrm{esc}$ does not lead to the requirement of higher values for $\delta$ to produce the LZ event.  For the SHM, the peak and width of the DM velocity distribution are set mainly by \vo.  As \vesc increases for a fixed value of \vo, $v_\mathrm{esc}$ will eventually reach a point where very few additional DM particles are added (even though they have higher speeds) and are no longer meaningfully contributing to the signal.  We thus see that for the $\delta\gtrsim566\,\mathrm{keV}$ ($\delta\gtrsim506\,\mathrm{keV}$), if $v_0<260\,\mathrm{km~s^{-1}}$ ($v_0<240\,\mathrm{km~s^{-1}}$) the SHM will not produce 1 event in the LZ energy range from $200~\mathrm{keV} - 300~\mathrm{keV}$, no matter how large \vesc is.  These \vo limits are relaxed down to $230~\mathrm{km~s^{-1}}$ and $215~\mathrm{km~s^{-1}}$ in the $N_{200-300}=0.0513$ benchmark. 

\textit{\textbf{Impact of the LMC}} -- Although the SHM is the canonical choice for the velocity distribution of DM, there is an extensive body of research that examines the limitations it might have in the context of the direct detection searches for DM (see Ref.~\cite{Green:2011bv, McCabe:2010zh} for reviews).  In particular, a recent study of $N$-body hydrodynamical simulations has indicated that the Large Magellanic Cloud (LMC) likely played a significant role in determining the high-speed tail of the DM velocity distribution in the solar neighborhood of the Milky Way~\cite{SmithOrlik2023LMC}.  We utilize the digitized version of the halo integrals from Figure 9 of Ref.~\cite{SmithOrlik2023LMC} to calculate the expected signals in LZ for the lower $1\sigma$, mean, and upper $1\sigma$ curves for Halo 13, which is the main focus in their study.  If the dark matter inelastic scattering cross section is velocity/momentum independent, the target recoil spectrum of Eq.~\ref{eq:dRdEr} will be proportional to this integral, often written as $\eta(v_\mathrm{min})$.  We find that only the lower $1\sigma$ curve can yield one event in the $200~\mathrm{keV}-300~\mathrm{keV}$ range within the bound of $\delta<505~\mathrm{keV}$,  with $N_{200-300}=1$ at $\delta=495~\mathrm{keV}$.  We find that all three curves fall below $N_{200-300}=0.0513$ at $\delta\sim 530~\mathrm{keV}$ and thus will produce no viable scenario to explain the LZ event for $\delta \gtrsim 566~\mathrm{keV}$.

\begin{figure}[t]
    \centering
    \includegraphics[width=\linewidth]{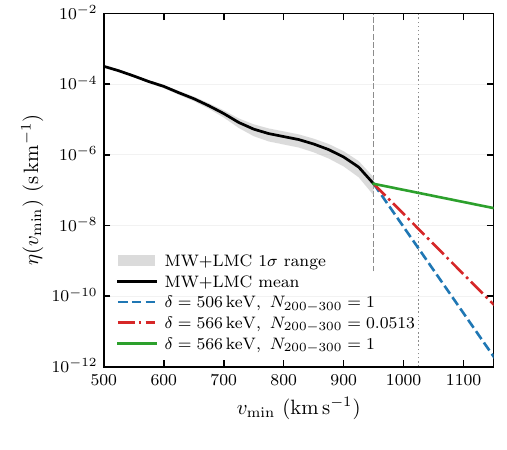}
    \caption{
    The halo integral, $\eta(v_{\min})$, from a recent study of $N$-body hydrodynamical simulations examining the impact of the Large Magellanic Cloud (LMC) on Milky Way's DM velocity distribution for $v_{\mathrm{min}}<950~\mathrm{km~s^{-1}}$ (dashed vertical line)~\cite{SmithOrlik2023LMC}.  The mean value is shown as a black curve and its $1\sigma$ range is shown as the gray shaded region. We explore exponential continuations beyond $950~\mathrm{km~s^{-1}}$ for the halo integral (see text for functional form). The dashed blue curve shows the tail required ($v_\mathrm{tail}=17.8~\mathrm{km~s}^{-1}$) for a mass splitting of $\delta=506~\mathrm{keV}$ to yield $N_{200-300}=1$.  The dot-dashed red curve ($v_\mathrm{tail}=25.4~\mathrm{km~s}^{-1}$) and the solid green curve ($v_\mathrm{tail}=126.3~\mathrm{km~s}^{-1}$) show the required extensions to yield $N_{200-300}=0.0513$ and  $N_{200-300}=1$, respectively, for the $\delta=566~\mathrm{keV}$ mass splitting. The dotted vertical line marks $v_{\min}\simeq1025~\mathrm{km\,s^{-1}}$, the approximate minimum speed required for a $248~\mathrm{keV}$ xenon recoil for a mass splitting of $\delta=566~\mathrm{keV}$.
    }
    \label{fig:mwlmc_tail_requirement}
\end{figure}

As the halo integral curves cut off at $\sim950~\mathrm{km~s^{-1}}$ in Ref.~\cite{SmithOrlik2023LMC}, we also examine a possible extension of $\eta(v_\mathrm{min})$ beyond this range to reproduce the LZ event.  In Figure~\ref{fig:mwlmc_tail_requirement}, we plot the digitized values of the mean halo integral as a black solid line, along with shaded bands corresponding to the halo integrals obtained from DM velocity distributions at $\pm1\sigma$ from the mean. We use a phenomenological extension of the mean halo integral by using the functional form 
\begin{equation}
    \eta(v_\mathrm{min})\sim\exp[-(v_{\mathrm{min}}-v_c)/v_\mathrm{tail}]
\end{equation}
where $v_c=950~\mathrm{km~s^{-1}}$ is the $v_\mathrm{min}$ value where the extension begins, and $v_\mathrm{tail}$ determines the slope of the extension.  For the $\delta=506~\mathrm{keV}$ benchmark, we find a very mild extension ($v_\mathrm{tail}=17.8~\mathrm{km~s}^{-1}$) that matches well the slope of the simulation at $v_\mathrm{min}=950~\mathrm{km\,s}^{-1}$, produces one event in the $200~\mathrm{keV}-300~\mathrm{keV}$ energy range. For the case of $\delta=566~\mathrm{keV}$, the dot-dashed red curve ($v_\mathrm{tail}=25.4~\mathrm{km~s}^{-1}$) and the solid green curve ($v_\mathrm{tail}=126.3~\mathrm{km~s}^{-1}$) show the required extensions to yield $N_{200-300}=0.0513$ and  $N_{200-300}=1$, respectively.  The slope of the red dashed curve is not significantly larger than slope of the simulation at $v_\mathrm{min}=950~\mathrm{km\,s}^{-1}$, whereas the green solid line is likely an unphysical scenario.  This indicates that both an upward fluctuation of the rate as well as this additional extension of the halo integral (or something similar) would likely be required to produce the LZ event for the $\delta=566~\mathrm{keV}$ mass splitting.  

\textit{\textbf{Conclusions}} -- The higgsino interpretation of the LZ event within the Standard Halo Model (SHM) is in significant tension with the solar-capture neutrino bounds of Ref.~\cite{Pospelov:2026ewn}. Notably, this tension remains even when the SHM parameters are pushed beyond the limits allowed by current measurements.  The $\delta\gtrsim566~\mathrm{keV}$ bound requires $v_0\gtrsim240~\mathrm{km~s^{-1}}$ in the SHM, independent of the Milky Way escape speed.  Significant tension remains even when adopting a dark matter velocity distribution motivated by recent hydrodynamical $N$-body simulations that incorporate the influence of the Large Magellanic Cloud. Under this distribution, one event in LZ's $200~\mathrm{keV}-300~\mathrm{keV}$ energy range is only possible for mass splittings up to $\delta=514~\mathrm{keV}$.

%
\textit{\textbf{Acknowledgements}} -- We thank Katherine Freese, Dionysios Theodosopoulos, and Patrick Stengel for helpful discussions. The work of A.\,G., I.\,C. and C.\,K.~is supported in part by the U.S.~Department of Energy, Office of Science, Office of High Energy Physics under Award Number DE-SC0024693. The work of C.\,K.~ is also supported by the U.S. National Science Foundation Growing Convergence Research award 2428507.


\bibliographystyle{apsrev4-1}
\bibliography{Refs}

\end{document}